\documentclass[10pt,twocolumn,letterpaper]{article}

\usepackage[pagenumbers]{cvpr}

\usepackage{amsmath,amssymb,amsthm}   
\usepackage{enumitem}

\usepackage{array}         
\usepackage{tabularx}

\usepackage{textcomp}   
\newcommand{\circlednum}[1]{\raisebox{0.8pt}{\textcircled{\raisebox{-0.9pt}{\scriptsize #1}}}}

\usepackage{siunitx}
\DeclareSIUnit{\byte}{B}   
\DeclareSIUnit{\inch}{in}  
\usepackage{tabularx}   
\usepackage{float}
\usepackage{siunitx}

\DeclareUnicodeCharacter{202F}{\,}  

\newcolumntype{Y}{>{\raggedleft\arraybackslash}X}
\theoremstyle{plain}
\newtheorem{theorem}{Theorem}

\newtheorem{lemma}{Lemma}

\theoremstyle{definition}

\definecolor{cvprblue}{rgb}{0.21,0.49,0.74}
\usepackage[pagebackref,breaklinks,colorlinks,allcolors=cvprblue]{hyperref}

\def\paperID{*****}
\def\confName{}
\def\confYear{}

\title{Sparse Graph Reconstruction and Seriation for Large-Scale Image Stacks}

\author{
\centering
\begin{tabular*}{\textwidth}{@{\extracolsep{\fill}} p{.30\textwidth} p{.30\textwidth} p{.30\textwidth}}
\centering\textbf{Fuming Yang}\\
\centering Harvard University\\
\centering {\footnotesize\hspace{0pt}\nolinkurl{fumingyang@fas.harvard.edu}}
&
\centering\textbf{Yaron Meirovitch}\\
\centering Harvard University\\
\centering {\footnotesize\hspace{0pt}\nolinkurl{yaron.mr@gmail.com}}
&
\centering\textbf{Jeff W.\ Lichtman}\\
\centering Harvard University\\
\centering {\footnotesize\hspace{0pt}\nolinkurl{jeff@mcb.harvard.edu}}
\end{tabular*}
}

\begin{document}
\maketitle

\begin{abstract}
We study the problem of recovering 1D order from noisy, local pairwise comparisons matrix while minimizing oracle queries. We cast this as reconstructing a sparse, noisy line graph and to our knowledge give the first algorithm that provably builds a sparse graph containing all edges needed for exact seriation using only \(O\!\bigl(N(\log N+K)\bigr)\) queries, near-linear in \(N\) for fixed window width \(K\). The method is parallelizable and applies to both binary and bounded-noise distance oracles. Our five-stage pipeline combines: (i) a random-hook Borůvka phase that connects components via short-range edges in \(O(N\log N)\) queries; (ii) iterative condensation to bound the graph diameter; (iii) a double-sweep BFS over the sparse graph to obtain a provisional global order; (iv) densification restricted to a \(K\)-window around this provisional sequence; and (v) a greedy \textsc{SuperChain} that assembles the final order. Under a simple top-1 margin and bounded relative noise, we prove exact recovery and show that \textsc{SuperChain} succeeds even when only about \(2N/3\) true adjacencies are present. We apply the approach to wafer-scale section ordering in serial section electron microscopy, which motivates the problem setting. Our method outperforms spectral, MST, and TSP baselines while requiring far fewer comparisons across diverse domains, and note that the algorithm is broadly applicable to sequencing tasks with local structure, including temporal snapshot ordering, archaeological seriation, playlist/tour construction, and other path-like data.
\newline 
Demo and code: \href{https://www.youtube.com/watch?v=f8raYfpha_k}{YouTube} \textbar\ \href{https://github.com/FumingYang-Felix/Sparse-Graph-Reconstruction-and-Seriation-for-Large-Scale-Image-Stacks}{GitHub}.
\end{abstract}

\sisetup{per-mode=symbol}
\DeclareSIUnit{\byte}{B}

\section{Motivation}
\label{sec:intro}

A complete map of synaptic connections is essential for developing mechanistic models of perception, learning, and neuropsychiatric disease. Over the past decades, connectomics has progressed from the \num{302}\,neurons \emph{C.\,elegans} map \cite{White1986Celegans}, to a fruit-fly brain containing more than 140 thousand neurons \cite{Dorkenwald2024FlyWire,Schlegel2024FlyWire}, to a petavoxel-scale (1\,mm$^3$) connectomes of the mouse visual cortex (MICrONS) \cite{MICrONSConsortium2025Nature} and human temporal cortex (H01) at nanometre resolution \cite{ShapsonCoe2024Petavoxel}.
The next major milestone, a synapse-level connectome of an entire adult mouse brain (500 mm³) is expected to generate exabytes of raw data. 

A typical connectomic pipeline starts with cutting resin-embedded brain tissue into thin sections (usually 30 nm), which are collected on a substrate and then imaged with EM. Section collection can be automated by using tape-collection, such as with the Automated Tape-collecting Ultramicrotome (ATUM \cite{hayworth2014wafermapper}), or by collecting sections directly on wafers (e.g. MagC \cite{Templier2019MagC}, Gauss-EM \cite{Fulton2024GaussEM}).

With the ATUM method, sections are collected in a sequence on a continuous tape, with the benefit that their order is preserved before EM begins. The tape is then cut into strips and mounted onto silicon wafers for imaging. While this method is robust, a single 100 mm silicon wafer will typically accommodate a relatively small number of sections (100-200 sections).

In contrast, tape-free methods can pack sections from hundreds to thousands per wafer. In MagC and Gauss-EM systems, make it more densely.
the sections are floated in a water bath and then pooled together over a silicon wafer. The water is withdrawn,  depositing the sections with high packing density on the wafer surface. 
These direct-to-wafer approaches have a number of advantages, including the above-mentioned high packing density, but also sections' flatness, reduced wrinkles owing to the annealing procedure, and the ability to collect very large sections, which make them more compatible with the requirements of imaging a whole-mouse brain with EM.  

Such large collections of images can be achieved using a IBEAM-mSEM pipeline in which semi-thin sections (250-500 nm) are subjected to multiple cycles of SEM imaging and ion-beam milling. \cite{hayworth2014wafermapper,Templier2019MagC,Fulton2024GaussEM}.   
However, a key distinction of these tape-free methods compared to tape-collecting methods, is that the original cutting section order is not preserved. Therefore, section order needs to be computationally recovered later to allow for volumetric alignment of the EM images \cite{hayworth2014wafermapper,Templier2019MagC}. This introduces several challenges due to the excessive number of large and highly variable sections.  

\begin{figure*}[t]
  \centering
  \includegraphics[width=\textwidth]{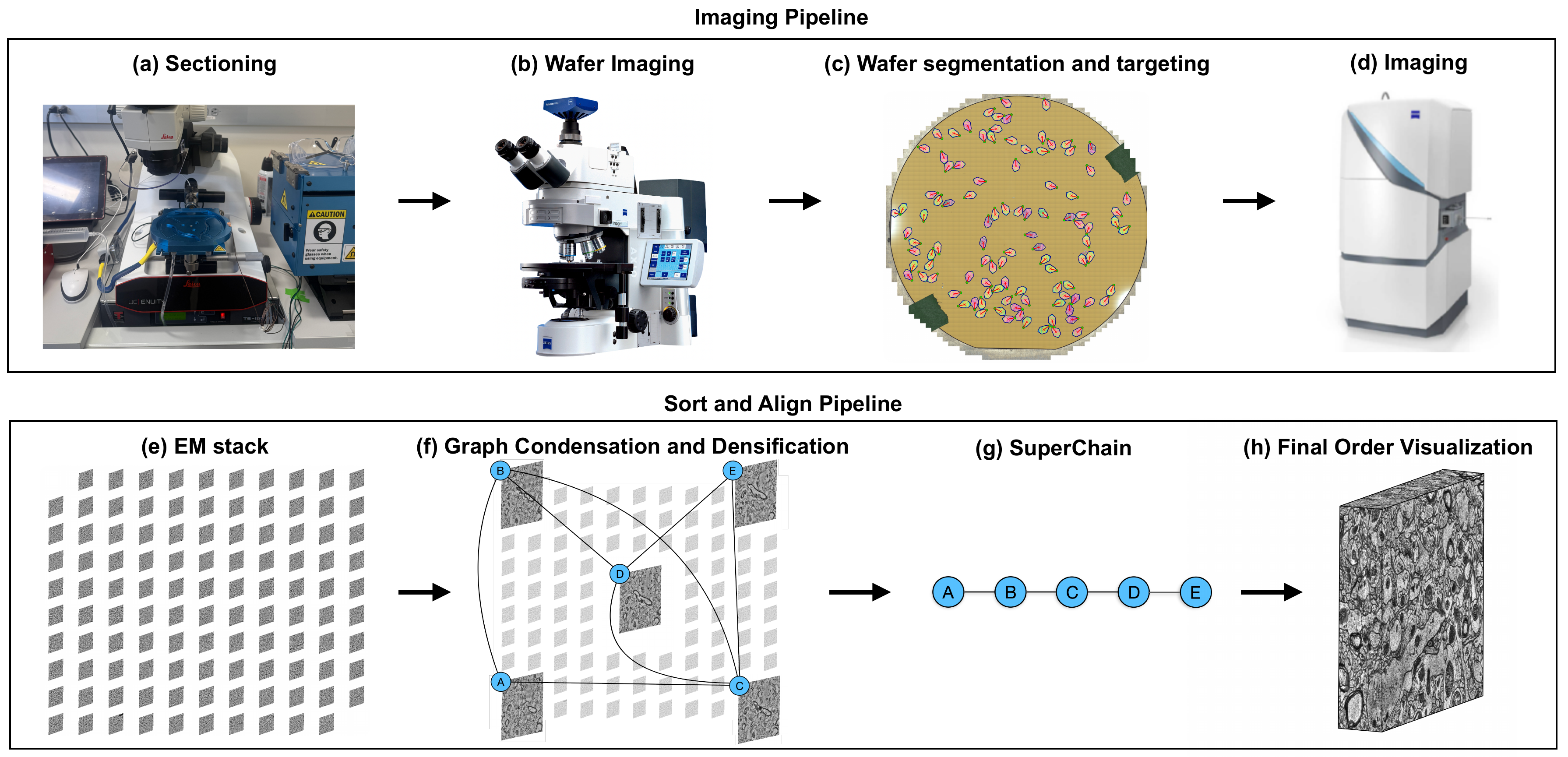}
  \caption{\textbf{WaferTools pipeline for wafer-to-volume EM sorting.}
  (a) Starting from unordered section collection method (e.g., MagC \cite{Templier2019MagC}), our WaferTools performs automatic section detection, (b–d) ROI unification with Procrustes alignment, and metadata generation for EM targeting. (e–g) The acquired EM stacks are then ordered via hybrid pipeline or near-linear Graph Condensation–Densification with SuperChain (hundreds to thousands sections). (f) Visualizing the final order result. Shown here is a wafer from the unpublished HI--MC dataset.}
  \label{fig:oracle_cost}
\end{figure*}

Current ordering methods fail to address three key challenges in EM section sorting: First, EM images usually yield a similarity pattern with a small set of strong matches limited to truly nearby sections, and a large noisy background elsewhere. As the distance between sections increases, pairwise scores based on intensity, SSIM, or SIFT inliers quickly drop to the noise level and fluctuate. In practice, since even nearby similarities are imperfect, incorporating noisy long-range comparisons does not resolve ambiguities but instead increases error rates \cite{hanslovsky2017,fogel2016serialrank}. 

Second, classic spectral seriation assumes that similarity decays monotonically with serial distance; the Fiedler vector of the graph Laplacian should therefore vary smoothly along the true sequence. However, when the leading band is flat and many off-diagonal blocks are missing at random, 
the second eigenvector can split into two nearly equal plateaux, and sorting its components produces block inversions \cite{Atkins1998Spectral,fogel2016serialrank,concas2023}. 

Third, comparing all image pairs is computationally expensive: identifying the few reliable short-distance edges would still require scoring a large fraction of the $N(N\!-\!1)/2$ pairs. Therefore, we seek a near-linear method that extracts the global order directly, avoiding near-exhaustive pair evaluation.



We address ordering under these constraints with a five-phase algorithm (Fig.~\ref{fig:oracle_cost}f,g). In the first phase, a binary graph is constructed using a modified random-hook Bor\r{u}vka procedure, which first connects the graph using \(\Theta(N\log N)\) binary oracle calls (each oracle call queries a single pairwise similarity). In the second phase, iterative farthest-point condensation reduces the diameter of the dynamically updated graph, bringing it near the diameter of the true order graph in \(\Theta(N\log N)\) steps. In the third phase, a double-sweep BFS over the sparse graph generates a provisional global seriation scaffold. In the fourth phase, we densify only inside a fixed window of \(K=\mathcal{O}(c)\) sections around this provisional sequence, adding edges along its backbone. Theory shows it suffices to add \(\Theta(NK)\) edges on that backbone for the recovery of missing true neighbors for each vertex even under considerable image noise (Sec.~\ref{sec:theory}). In the final phase, a greedy routine dubbed \textsc{SuperChain} exploits the empirical gap between the strongest edge at each vertex and the next-best alternative to assemble the true order permutation in \(\Theta(N\log N)\) time. The overall budget \(\Theta\!\bigl(N(\log N+K)\bigr)\) is near-linear, reducing the one-million-section problem to a tractable overnight job on a single workstation.

In summary, in this work we formulate the EM seriation problem using concepts from random graph theory. To our knowledge, it introduces the first near-linear-query algorithm for the seriation problem. Experiments confirm zero ordering error and an eight-fold speedup over dense spectral methods. Finally, to aid reproducibility and further work in this direction, we release a \SI{500}{sections} benchmark dataset with open-source code.

\subsection{Related Work}
\label{sec:related}

\noindent
We use \emph{seriation} to denote the algorithmic problem of recovering a linear
order from pairwise similarities (a sparse, partially observed graph in our case),
and \emph{ordering} to refer to the task and its output permutation.

Early algorithms for recovering section order leveraged image-similarity matrices
(e.g., Pearson correlation or feature matches) together with manual curation; subsequent work
formalized \emph{post–acquisition} correction and jointly estimated section order and z–spacing
directly from image-derived similarities \cite{hanslovsky2017}. 

Two common approaches employ Minimum Spanning Tree (MST) and Traveling Salesman Problem (TSP) formulations. 
MST-based methods build a minimum spanning tree on the similarity (or distance) graph (in polynomial time once the graph is known; see \citet{Kruskal1956,Prim1957}) and then derive a sequence from a traversal; they work when reliable short-range edges are present, but single-linkage–style chaining means a few wrong links can propagate errors \citep{SneathSokal1973}. 
TSP-style approaches seek a minimum-cost Hamiltonian path over sections; exact TSP is NP-hard \citep{Karp1972}, and practical solvers typically rely on dense costs or carefully curated candidate edges \citep{LinKernighan1973,Helsgaun2000,Applegate2006}. 
In our data, the informative signal is highly local: the nearest neighbor of a section is typically a true neighbor, the second true neighbor often appears among the next few best matches, and similarities for offsets $\ge\!10$ are at the noise level. 
Consequently, MST traversals and TSP heuristics either require scoring a large fraction of pairs to be robust, or they risk local swaps and occasional incorrect links.

In practice, spectral seriation and related convex relaxations (e.g., 2–SUM/SerialRank) provide more principled baselines
\cite{Atkins1998Spectral,fogel2016serialrank}. Spectral seriation sorts the components of the Fiedler vector of the graph Laplacian; however,
when the leading similarity band flattens or the Laplacian admits a \emph{multiple} Fiedler value,
the solution becomes non-unique and can induce block inversions/``zig–zag'' artifacts
\cite{Atkins1998Spectral,concas2023,fogel2016serialrank}.


The seriation problem considered here appears in other domains, including archaeology and computational biology.
Classical spectral methods \cite{Atkins1998Spectral} and modern convex/spectral formulations such as
SerialRank \cite{fogel2016serialrank} provide robustness to noise and missing entries under appropriate models,
but many pipelines still rely on a large number of pairwise measurements, which are prohibitive at 10\,mm$^3$ scale.



\section{Methodology}
\label{sec:methodology}

Our approach to the section ordering problem consists of two main components: first, we construct a sparse graph that captures the essential connectivity structure while avoiding quadratic complexity, and second, we recover the exact ordering from this sparse graph. We begin by describing the overall pipeline before providing rigorous theoretical analysis.

\subsection{Sparse Graph Construction}
\label{sec:sparse_construction}

The fundamental challenge in section ordering is that exhaustive pairwise comparison of $N$ sections requires $O(N^2)$ similarity computations, which becomes prohibitive for large-scale connectomics projects. We address this by constructing a sparse graph containing only $O(N(\log N + K))$ edges that provably includes all connections necessary for exact seriation, where $K$ is a small constant densification window.

\subsubsection{Feature Extraction and Similarity Computation}

For each candidate section pair, we implement a multi-stage feature-matching pipeline to determine edge existence and weight. We first extract SIFT keypoints from both sections and establish putative correspondences via FLANN-based nearest neighbor matching. These matches are then filtered using RANSAC-based geometric consistency checks, where correspondences inconsistent with an affine transformation model are rejected as outliers. For pairs passing the minimum inlier threshold, we compute the structural similarity index (SSIM) on the overlapping region after alignment, capturing fine-scale morphological consistency beyond keypoint matching. The final similarity score combines both metrics:
\begin{equation}
M_{uv} = \mathrm{SSIM}(S_u, S_v) \cdot \#\text{Inliers}(S_u, S_v)
\label{eq:similarity_score}
\end{equation}
This multi-modal scoring ensures robust edge detection despite imaging artifacts and registration errors.

\begin{figure*}[t]
    \centering
    \includegraphics[width=\textwidth]{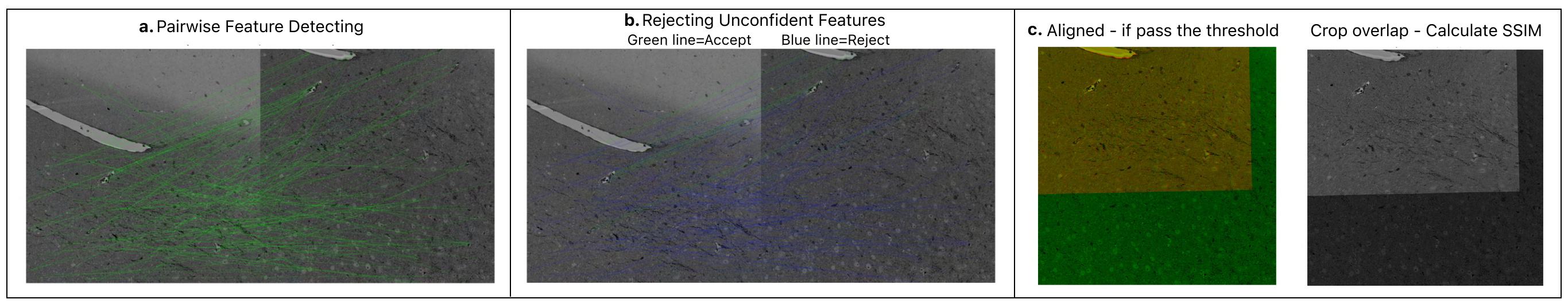}
    \caption{\textbf{Feature-matching pipeline for similarity computation.} (a) SIFT keypoints are extracted and matched between candidate section pairs using FLANN. (b) RANSAC geometric verification filters matches to retain only those consistent with an affine transformation, rejecting outliers. (c) For pairs exceeding the inlier threshold, sections are aligned and SSIM is computed on the overlap region to capture fine morphological similarity.}
    \label{fig:feature_pipeline}
\end{figure*}

\subsubsection{Four-Phase Graph Construction Algorithm}

Our sparse graph construction proceeds through four carefully designed phases that balance computational efficiency with theoretical guarantees. The key insight is that we can build a graph containing all essential edges without examining most of the $N(N-1)/2$ possible pairs.

\paragraph{Phase 1: Initial Connectivity via Random-Hook Borůvka.}
We begin by establishing global connectivity through a randomized variant of Borůvka's algorithm. Unlike classical MST algorithms that assume access to all edges, our setting requires discovering which edges exist through oracle queries. Each connected component samples $O(\log N)$ random candidate partners per round and queries the oracle for valid short-range connections. When a valid edge is found, the components are marked for merging. This process continues until a single spanning tree is formed, requiring $O(N\log N)$ queries with high probability under the assumption that short edges between components exist and can be found through random sampling. Under the assumptions in Sec.~\ref{sec:theory} with a constant $K$, Phase~A needs $O(N\log N)$ probes with high probability; if no truly short cross-component edge passes the acceptance test (e.g., due to an overly noisy oracle or undersized candidate pool), Bor\r{u}vka cannot merge components and may stall, in which case our guarantees do not apply.

\begin{figure}[t]
    \centering
    \includegraphics[width=1.05\linewidth]{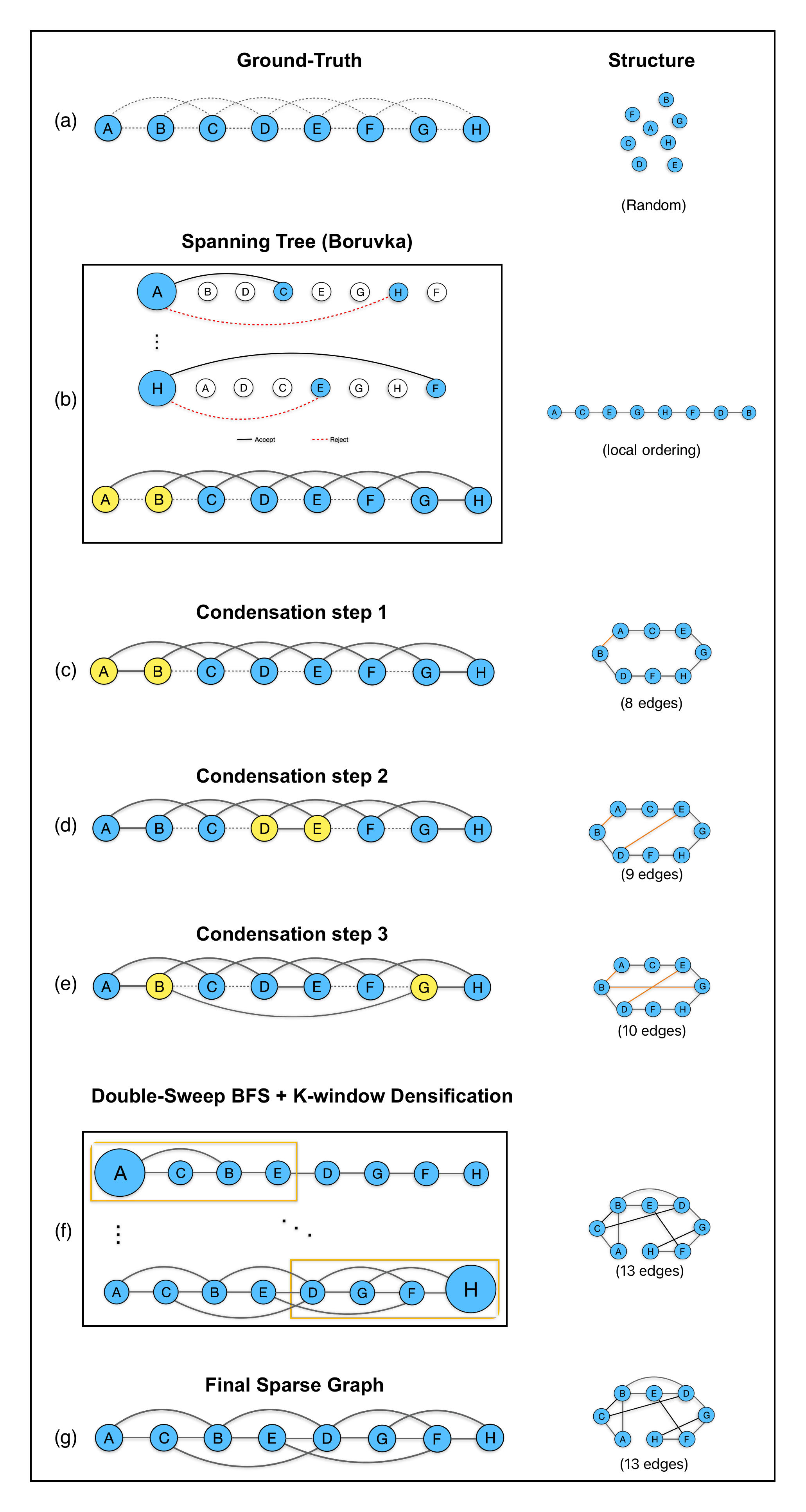}
    \caption{\textbf{Four-phase sparse graph construction.} (a) Ground-truth linear structure to be recovered. (b) Phase 1: Random-hook Borůvka builds initial spanning tree using $O(N\log N)$ queries. (c-e) Phase 2: Iterative condensation adds long-range edges to reduce graph diameter. (f) Phase 3: Double-sweep BFS produces provisional ordering; Phase 4: $K$-window densification recovers missing local edges. (g) Final sparse graph containing all essential edges for exact seriation.}
    \label{fig:pipeline}
\end{figure}

\paragraph{Phase 2: Diameter Reduction through Iterative Condensation.}
The spanning tree from Phase 1 ensures connectivity but may have large diameter, potentially placing true neighbors far apart in the tree metric. We address this through iterative condensation, which systematically reduces the graph diameter by adding carefully selected edges. In each round, we identify vertices farthest from a landmark set (initially the diameter endpoints) and query for all their short-range connections. These new edges create shortcuts that provably halve the effective diameter. After $O(\log N)$ rounds, the graph diameter is reduced to $O(c)$, where $c$ is the short-range radius, ensuring that the subsequent BFS ordering places true neighbors within a bounded window.

\paragraph{Phase 3: Global Ordering via Double-Sweep BFS.}
With the graph diameter bounded, we establish a provisional global ordering using double-sweep breadth-first search. Starting from an arbitrary vertex, we identify the farthest vertex, then find the vertex farthest from that, establishing the graph's diameter endpoints. All vertices are then ordered by their BFS distance from one endpoint. While this ordering may not be perfect, the bounded diameter guarantee ensures that any two adjacent sections in the true order appear within $O(c)$ positions of each other in this provisional arrangement.

\paragraph{Phase 4: Local Densification within Fixed Windows.}
The final phase ensures no essential edges are missed by querying all pairs within a fixed window $K$ around each vertex's position in the provisional ordering. Specifically, for each vertex at position $i$, we query all vertices at positions $j$ where $|i-j| \leq K$. With $K = O(c)$, this guarantees that all true adjacent pairs are queried, while the total number of queries remains $O(NK) = O(N)$. The resulting graph contains all edges necessary for exact recovery while maintaining sparsity.

\subsection{Sorting on the Sparse Graph}
\label{sec:sorting_sparse}

Given the sparse graph from the construction phase, we need to extract the unique linear ordering. The challenge is that even with all true edges present, the graph may contain spurious connections and cycles that complicate direct ordering methods. We address this through a two-stage approach: first building an Iterative Similarity Search (ISS) graph that retains only the most reliable connections, then applying our SuperChain algorithm to assemble the final ordering.

\subsubsection{Iterative Similarity Search Graph Construction}

From the sparse similarity graph, we construct a more selective ISS graph where each vertex retains only its two strongest connections. Formally, for each section $S_i$, we identify:
\begin{align}
j_1 &= \arg\max_{j \neq i} M_{ij} \\
j_2 &= \arg\max_{j \neq i, j \neq j_1} M_{ij}
\end{align}
The ISS graph $G_{\mathrm{ISS}} = (V, E_{\mathrm{ISS}})$ contains edges $E_{\mathrm{ISS}} = \{\{i,j_1\}, \{i,j_2\} \mid i = 1,\ldots,N\}$, resulting in at most $2N$ edges. Vertices with a degree greater than 2 are identified as hubs and temporarily removed, decomposing the graph into simple paths and cycles that serve as building blocks for the final assembly.

\subsubsection{SuperChain Assembly Algorithm}

The SuperChain algorithm reconstructs the global ordering by iteratively merging the path components according to four carefully designed rules that ensure correctness:

\begin{enumerate}[label=\textbf{R\arabic*},leftmargin=2em]
\item \textbf{Maximize Chain Rule} — In any round, if the same vertex is selected multiple times, keep only the candidate with the highest similarity score.
\item \textbf{Unbroken Chain Rule}. Once a vertex becomes internal to a chain (having neighbors on both sides), it cannot acquire additional connections, preserving the linear structure.
\item \textbf{End-to-End Chain Rule}. Chains can only be joined at their endpoints, maintaining consistent orientation throughout the merging process.
\item \textbf{Fallback Chain Rule}. When direct endpoint connections are unavailable, we allow examining edges from vertices one step inward from the endpoints, providing robustness to local errors.
\end{enumerate}

\begin{figure}[t]
    \centering
    \includegraphics[width=1.0\linewidth]{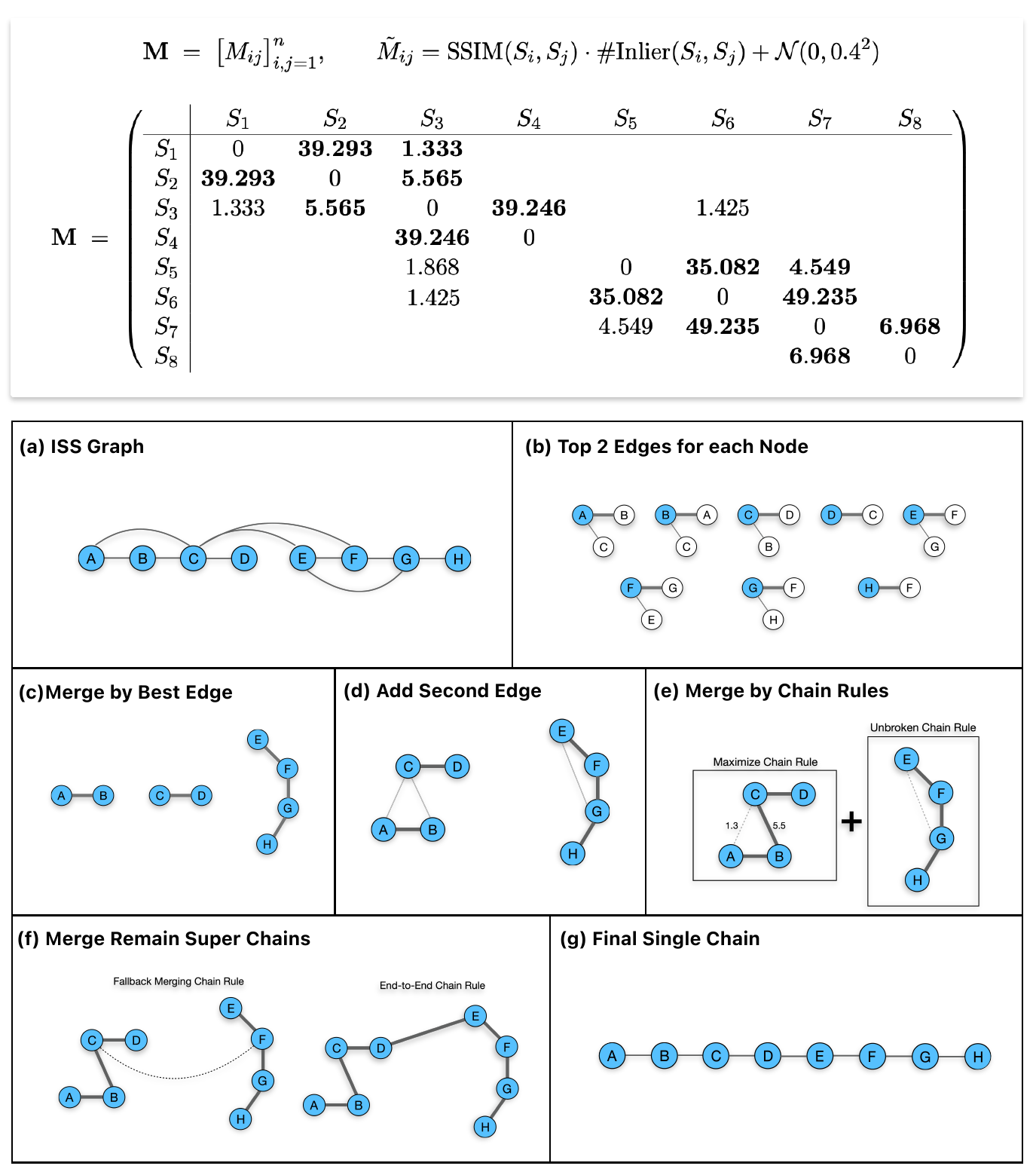}
    \caption{\textbf{SuperChain assembly process.} (a) Initial ISS graph with each vertex retaining its two strongest edges. (b-c) Stage 1: Reciprocal best matches form mini-chains. (d) Stage 2: Chains extend via endpoint connections. (e) Chain rules ensure correct assembly: maximize-chain prevents wasteful merges, unbroken-chain maintains linearity. (f) Fallback rule allows depth-1 connections when needed. (g) Final single  representing the complete section order.}
    \label{fig:superchain}
\end{figure}

The algorithm proceeds through four stages. First, depth-first traversal extracts initial mini-chains from the ISS graph. Second, chains are iteratively merged using their strongest inter-chain edges, with a max-heap ensuring globally optimal decisions. Third, any remaining endpoint connections are exploited for additional merges. Finally, the fallback rule is applied to handle residual fragmentation (current endpoints do not have similarity scores in the sparse matrix). Throughout this process, the chain rules guarantee that the true ordering is preserved whenever the 2N/3 true edges are existing.

\section{Experiments}

\begin{figure}[t]
    \centering
    \includegraphics[width=1.0\linewidth]{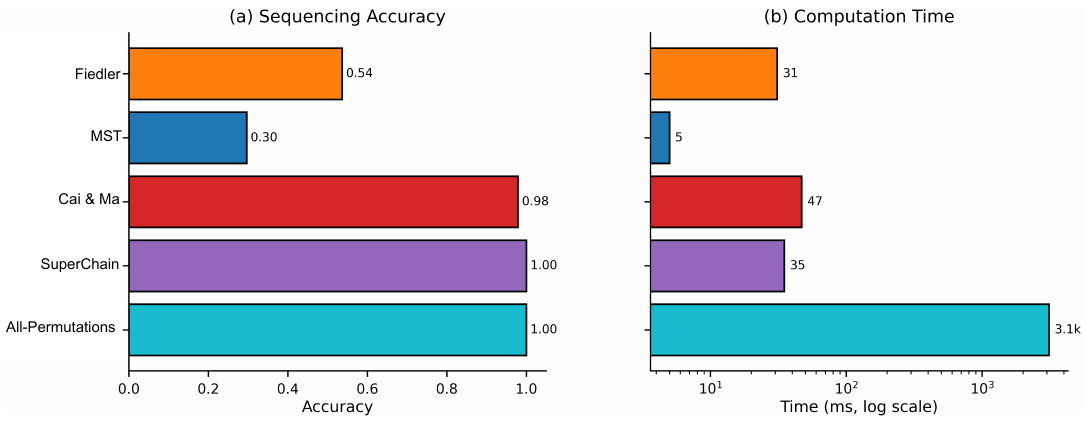}
\caption{\textbf{Comparison of sequencing algorithms on an 8-section H01 example.}
\textbf{(a)} Accuracy (Eq.~\ref{eq:acc}) measures order fidelity (higher is better).
\textbf{(b)} Wall-clock time (ms) above each bar reports the \emph{ordering-only} stage for each method.
“Fiedler” denotes spectral seriation via the second eigenvector of the normalized Laplacian (on the dense similarity matrix)\citep{Atkins1998Spectral,Fiedler1973};
“Naive MST” builds a minimum spanning tree on the dense graph and linearizes it by traversal;
“Cai \& Ma (2024)” refers to the adaptive matrix-reordering algorithm of \citet{CaiMa2024MatrixReorderingTIT}.
\textsc{SuperChain} attains near-perfect accuracy with substantially lower runtime, illustrating the practical complexity gap.
}
\label{fig:superchain-comparison}
\end{figure}

We evaluate our proposed seriation framework along five axes. 
(1) Scalability: we compare similarity-only (the cost of building pairwise similarity matrix, Fig.~\ref{fig:CondensationTime}), ordering-only (Fig.~\ref{fig:superchain-comparison}(b),~\ref{fig:superchain-comparison2}(b)); and full end-to-end runtime as the number of sections increases is reported in Tab.~\ref{tab:runtime-marked}. 
(2) Accuracy: accuracy of the recovered order is shown on the eight sections H01 example in Fig.~\ref{fig:superchain-comparison}(a) and across larger volumes (1000 sections) in Fig.~\ref{fig:superchain-comparison2}(a); these panels expose how baseline methods degrade while SuperChain maintains high order accuracy. 
(3) Empirical validity of the margin assumptions: statistics of the observed $\Delta$-margin and its distribution are summarized in Tab.~\ref{tab:delta}, validating the assumptions used in the theoretical analysis. 
(4) Worst-case scenario analysis: Monte Carlo trials under adversarial fragmentation showing that SuperChain achieves reliable exact recovery even when only about $2N/3$ internal sections have a correct second-best neighbour (Fig.~\ref{fig:recovery_probability}).
(5) Component contributions: ablations isolating condensation, densification window size, and hook variants appear in Tab.~\ref{tab:ablation}, quantifying each stage's effect on edge-edit rate and cost.

For evaluation and model selection, we need a single scalar that ranks competing variants/baselines by how close their orderings are to the target; we therefore define a cost that measures the distance to the true order.
Given a candidate graph $\hat G=(V,\hat E)$ produced by a variant, we measure its distance to $G^\star=(V,E^\star)$ via the edge-edit distance (the symmetric difference of edge sets):
\begin{equation}
  \operatorname{EED}(\hat E,E^\star)
  \;=\;
  |\hat E\triangle E^\star|
  \;=\;
  |\hat E\setminus E^\star|+|E^\star\setminus\hat E|
  \label{eq:eed}
\end{equation}
Because the entire wafer can be reversed, we score the
smaller of the two directions,
\begin{subequations}
\begin{equation}
\operatorname{Cost}_{\mathrm{edge}}
=\min\left\{
  \operatorname{EED}(\hat E, E^\star),
  \operatorname{EED}(\hat E, E^{\star R})
\right\}
\end{equation}
\begin{equation}
\begin{aligned}
\text{Accuracy}&=\;
  1-\min\!\bigl\{\operatorname{EED}(\hat E,E^\star),
               \operatorname{EED}(\hat E,E^{\star R})\bigr\}
\end{aligned}
\label{eq:acc}
\end{equation}
\end{subequations}

\subsection{Scalability and Accuracy}\label{sec:noise-exp}
\begin{figure}[t]
    \centering
    \includegraphics[width=1.0\linewidth]{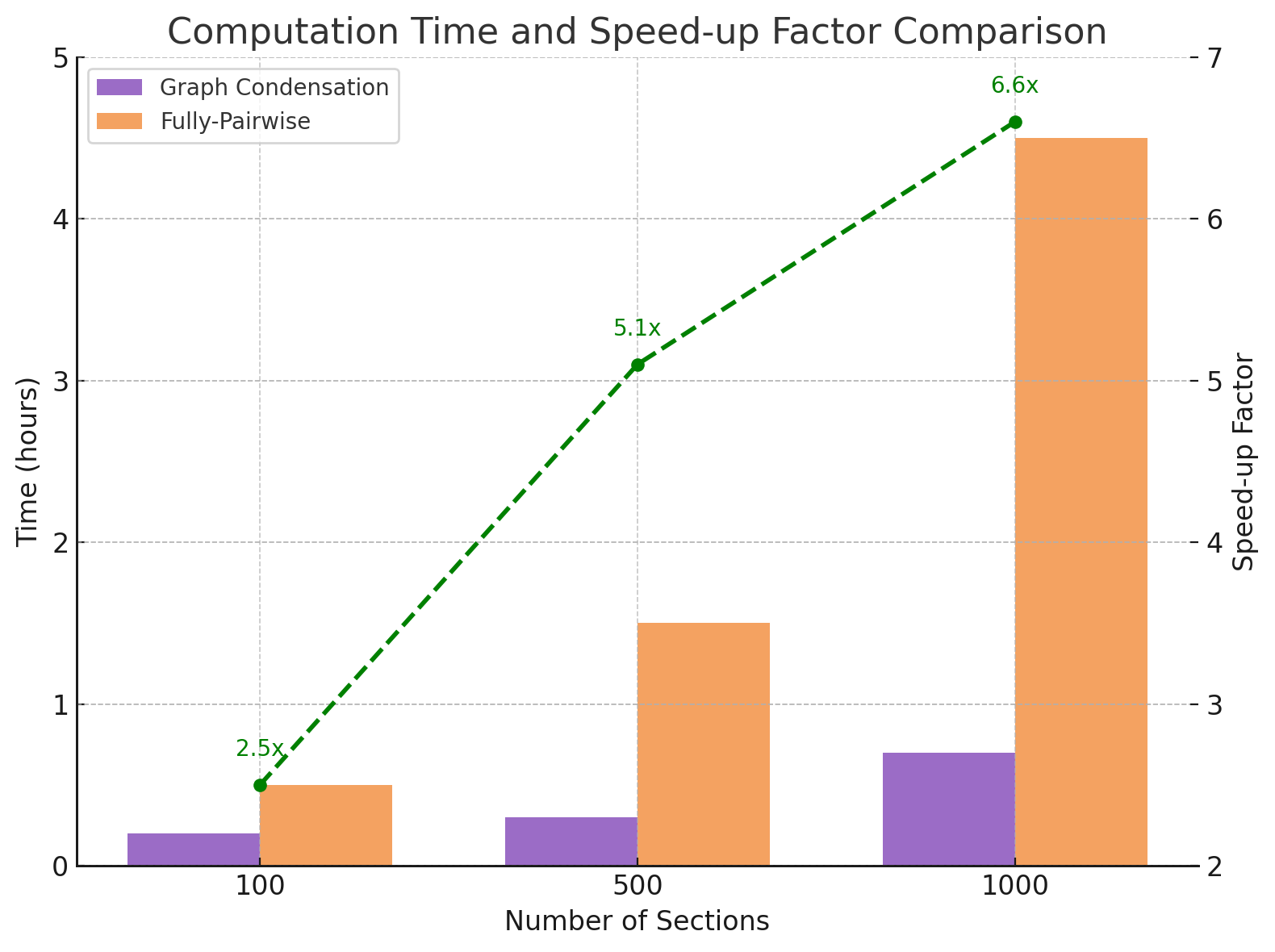}
    \caption{\textbf{Similarity Computation Time and Speed-up Factor Comparison.} 
We compare the runtime (bar plot, left y-axis) and speed-up factor (dashed line, right y-axis) of our proposed Graph Condensation approach versus the traditional fully-pairwise method across 100, 500, and 1000 EM sections. 
}
    \label{fig:CondensationTime}
\end{figure}

We compared the computational efficiency of our proposed Graph Condensation strategy with a conventional fully-pairwise alignment approach across 100, 500, and 1000 sections (Fig.~\ref{fig:CondensationTime}). On the 100-section dataset, Graph Condensation already achieves a 2.5× speed-up, reducing computation time from approximately 0.2 hours to less than 0.1 hours. The efficiency gain becomes more pronounced as dataset size increases: at 500 sections, the method achieves a 5.1× speed-up, and at 1000 sections, a 6.6× speed-up, lowering the runtime from more than 4.5 hours to under 1 hour.
These results demonstrate that while high-resolution fully-pairwise computations scale quadratically with the number of sections, Graph Condensation grows much more gently, approaching near-linear complexity in practice. Crucially, this efficiency gain does not compromise accuracy, since the condensation framework preserves the most informative edges through Borůvka contraction and densification while discarding redundant comparisons. As a result, Graph Condensation enables wafer-scale ordering of thousands of sections that would otherwise be computationally costly under fully-pairwise schemes.

Our similarity matrices exhibit strong block structure and large dynamic-range edges, violating standard assumptions for spectral methods. This explains the observed position swaps in spectral ordering.

We benchmarked four representative sequencing algorithms on H01 datasets ranging from 8 to 1000 sections (33nm thickness) and a separate 100-section dataset (250nm thickness). Na\"{i}ve MST and Fiedler methods exhibited poor accuracy (0.30 and 0.54, respectively), despite offering low runtime, highlighting their inability to preserve correct global order under realistic similarity distributions. The Cai \& Ma (2024) method significantly improved accuracy (0.98) but at the cost of substantially higher runtime. In contrast, SuperChain achieved perfect accuracy (1.00) with runtime only marginally higher than MST and Fiedler, demonstrating a superior balance between accuracy and speed.

For comparison, the exhaustive All-Permutations baseline also attained perfect accuracy but required orders of magnitude more computation time, 
rendering it impractical for real-world use.

Together, these results emphasize that SuperChain uniquely combines state-of-the-art accuracy with near-optimal runtime, enabling tractable and reliable ordering for both small and large EM datasets.

\begin{figure}[H]
    \centering
    \includegraphics[width=1.0\linewidth]{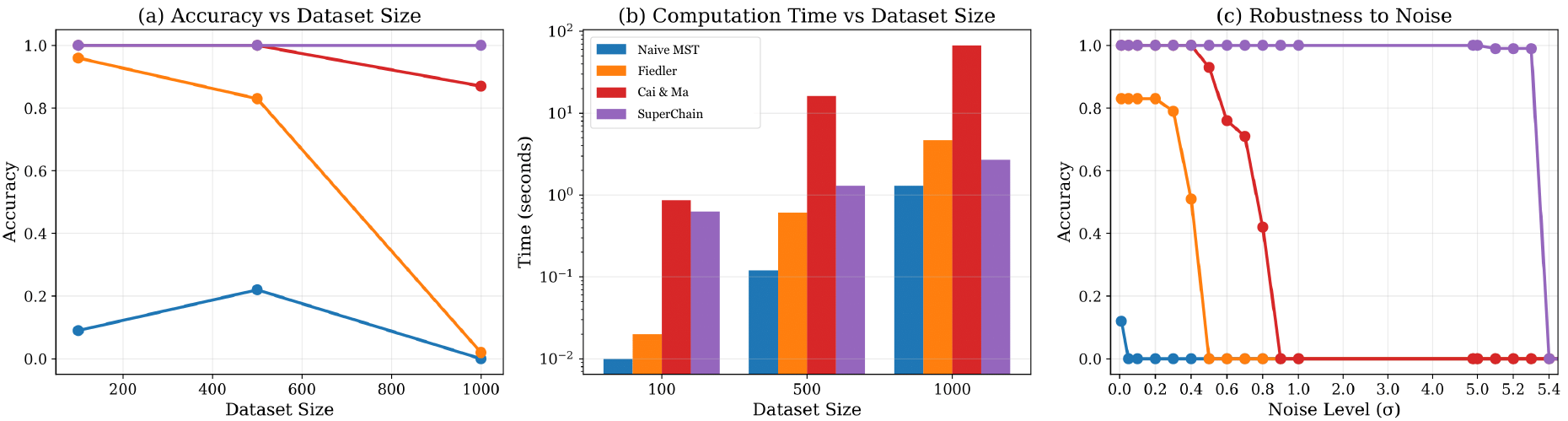}
    \caption{\textbf{Accuracy and Computation Time Comparison.} {
\textbf{(a)} The accuracy across 100, 500, 1000 sections EM image stacks - H01 (Public) dataset. 
\textbf{(b)} Computational time among all four methods.
\textbf{(c)} Robustness of each method by adding Gaussian noise}
}
    \label{fig:superchain-comparison2}
\end{figure}

To evaluate the performance of our proposed seriation framework across varying dataset sizes and noise conditions, we conducted experiments on the public H01 human cortex dataset (100, 500 and 1000 33nm thick sections). The results are summarized in Fig.~\ref{fig:superchain-comparison2}.

While Naïve MST and Fiedler show a marked decline in accuracy as the number of sections increases, SuperChain consistently maintains perfect accuracy across all dataset sizes. The Cai \& Ma (2024) method performs competitively on smaller stacks but begins to degrade at 1000 sections. Notably, on the 100-semi-thin section dataset, both Cai \& Ma (2024) and SuperChain achieve accuracies above 0.95, whereas MST and Fiedler remain below 0.3 and 0.9, respectively.

Runtime analysis highlights the scalability advantage of SuperChain. On the 
1000-section H01 dataset, SuperChain completes ordering in under 10 seconds, 
while Cai \& Ma (2024) requires more time. Fiedler and MST show better time complexity but at the cost of accuracy, making them unsuitable for large-scale reconstructions.

\begin{table}[ht]
  \centering
  \footnotesize 
  \caption{
    Overall end-to-end runtime comparison (minutes), including similarity computation and sequencing, on 100, 500, and 1000 EM sections. 
    All timings are wall-clock measured on a single CPU core.
    SuperChain’s ordering-only time is under 10 seconds for 1000 sections (Fig.~\ref{fig:superchain-comparison2}); the numbers shown here are full-pipeline. 
  }
  \setlength{\tabcolsep}{5pt}
  \begin{tabular}{lccc@{\hspace{0.8em}}c}
    \toprule
    \textbf{Method} & \textbf{100 Sections} & \textbf{500 Sections} & \textbf{1000 Sections} & \\[-0.4em]
    \midrule
    Naive MST       & 24.91 &  90.12 & 271.3 & \circlednum{1} \\[-0.3em]
    Fiedler         & 24.92 &  90.61 & 274.7 & \circlednum{1} \\[-0.3em]
    Cai\&Ma~2023   & 25.77 & 106.30 & 337.3 & \circlednum{1} \\[-0.3em]
    SuperChain      & 10.13 &  19.00 &  43.9 & \circlednum{2} \\
    \bottomrule
  \end{tabular}
  \vspace{4pt}

  \raggedright
  \circlednum{1} Full pairwise similarity + algorithm-specific ordering. \\
  \circlednum{2} Graph condensation-densification + SuperChain (full pipeline ordering; ordering-only time is much lower). 
  \label{tab:runtime-marked}
\end{table}

Together, these results establish that SuperChain is uniquely capable of combining scalability, accuracy, and robustness, enabling reliable section ordering of both ultrathin (30-40 nm) and semi-thin (250 nm) sections.

Tab. ~\ref{tab:runtime-marked} demonstrates a clear scalability gap between the full–pairwise pipeline, and our condensation–densification pipeline.  
Naïve MST, Fiedler ordering, and the Cai \& Ma (2024) variant all rely on dense similarity matrices; their runtime grows roughly quadratically, reaching 4.5--5.5 hours for 1\,000 sections. In contrast, SuperChain---which first sparsifies the graph via condensation and then locally densifies it with a $K$-window scheme remains consistently faster: 2.5$\times$ at 100 sections, $\sim$5$\times$ at 500 sections, and 6--8$\times$ at 1\,000 sections (43.9 min vs.\ 271--337 min). These results highlight that reducing the edge set before seriation is critical for connectomics datasets - where thousands of semithin slices must be ordered under tight time budgets.

\subsection{Empirical $\Delta$-margin}
\label{sec:delta}

For every section, we compute the edge–weight gap.
\begin{equation}
\begin{aligned}
A_i &:= \max_{u\not\in\{v_{i-1},v_{i+1}\}} w(v_i,u),\\
\Delta(v_i) &= w(v_i,v_{i-1}) - A_i,\quad
\Delta = \min_i \Delta(v_i).
\end{aligned}
\label{eq:delta}
\end{equation}

Tab.~\ref{tab:delta} reports the resulting statistics.
All datasets exhibit $\Delta\ge 0.09$, safely above the
theoretical bound
$1/c-1/\tau\approx0.5/c\;(=0.05)$ derived in
Sec.~\ref{sec:superchain}, thus validating Assumptions~A1–A3.

\begin{table}[t]
  \footnotesize
  \setlength{\tabcolsep}{6pt}
  \centering
  \caption{Empirical $\Delta$-margin across datasets.}
  \vspace{0.3em}
  \begin{tabular}{lccc}
    \toprule
    Dataset & $N$ & $\min\Delta$ & $\text{mean}\,\Delta\pm\text{std}$ \\
    \midrule
    HPF (250 nm) & 100  & 0.23 & $0.27\pm0.07$ \\
    H01           & 500  & 0.10 & $0.13\pm0.03$ \\
    H01           & 1000 & 0.09 & $0.11\pm0.02$ \\
    \bottomrule
  \end{tabular}
  \label{tab:delta}
  \vspace{-0.7em}
\end{table}

\subsection{Numerical Validation under Worst-Case Fragmentation}\label{sec:montecarlo}

To validate Theorem~\ref{sec:superchain}, 
we designed a Monte Carlo experiment on synthetic chains of length $N=12$ 
constructed under the worst-case fragmentation scenario implied by Assumption~A1. 
Specifically, we ensured that every section’s top-1 neighbour was correct, 
but arranged them so that the initial best-pair graph decomposed into 
six disjoint mini-chains of length two: 
\[
[1,2], [3,4], [5,6], [7,8], [9,10], [11,12].
\]
This constitutes the hardest possible case, as without additional evidence 
the true chain cannot be uniquely reconstructed.

We then progressively introduced second-best correctness. 
For each trial, a fixed proportion $p\in\{0\%,20\%,40\%,60\%,80\%,100\%\}$ 
of the $N-2$ internal sections were randomly assigned a second-best edge 
consistent with their true neighbour.  
For the remaining sections, the second-best was replaced by a plausible 
but false candidate chosen uniformly from a $\Delta\!=\!4$ neighbourhood, 
excluding the true neighbours.  
This simulates realistic ambiguity: nearby non-adjacent sections 
often exhibit non-trivial similarity.

\begin{figure}[t]
  \centering
  \includegraphics[width=1.0\linewidth]{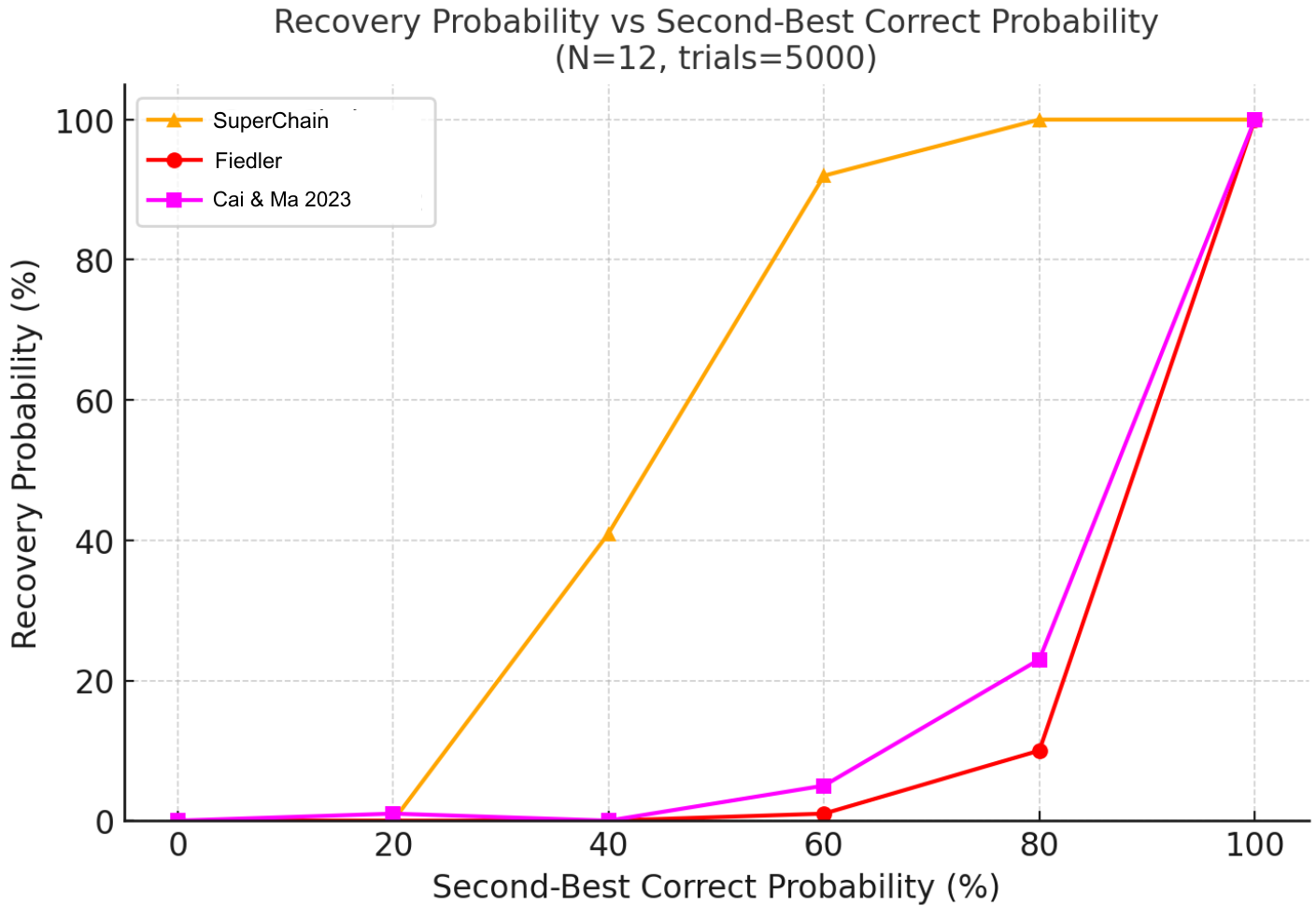}
  \caption{\textbf{Monte Carlo recovery probability of the 
  second-best edge correctness rate.}
  The analysis assumes the best edges are 100\% correct, ensuring 
  reliable local connections, but considers the worst-case regime
  where these best edges form the maximum number of disjoint mini-chains 
  ($N/2=6$ for $N=12$). Recovery therefore depends critically on the 
  correctness of second-best edges, each correct with probability $p$. 
  Results are averaged over $5{,}000$ trials with $N=12$. 
  SuperChain demonstrates markedly higher robustness, maintaining over 
  90\% recovery probability even at $p=60\%$.}
  \label{fig:recovery_probability}
\end{figure}

Each configuration was sampled over 5000 Monte Carlo trials. 
Recovery was counted as successful if the output permutation 
matched the ground-truth order $(1,2,\dots,12)$ or its reversal (Fig.~\ref{fig:recovery_probability}).

The results confirm our theoretical analysis.  
Fiedler ordering and Cai \& Ma(2024) essentially collapse to random 
performance unless second-best correctness is perfect ($100\%$).  
In contrast, SuperChain retains high recovery rates even when only 
$\sim 2N/3$ sections have a correct second-best neighbour.

\subsection{Ablation Study}
\label{sec:ablation}

We evaluated component contributions using H01 subsets (\(N\!=\!500,\;1000,\;5000\) sections) with added Gaussian noise\(\varepsilon=0.05\). Results are averaged over three trials.

\paragraph{Variants.}
No-C: skip Phase (P2) condensation.  
No-D: skip \(K\)-window densification.  
Small-K: set \(K=c\).  
Large-K: set \(K=4c\).  
Rand-Hook: in Borůvka, each component probes one neighbour per
round instead of \(\Theta(\log N)\).  
Ours: full pipeline.

\begin{table}[t]
\footnotesize
\setlength{\tabcolsep}{4pt}
\centering
\caption{Ablation on H01 subsets ($\varepsilon=0.05$).  
Edge-edit rate ↓ is better; calls/$N$ measures oracle efficiency.}
\vspace{0.3em}
\begin{tabularx}{\linewidth}{@{}lYYYYY@{}} 
\toprule
 & \multicolumn{3}{c}{\textbf{edge-edit rate} (\%)} &
   \textbf{calls/$N$} & \textbf{time (s)}\\
\cmidrule(lr){2-4}
Variant & 500 & 1k & 5k & calls/$N$ & time (s)\\ \midrule
No-C        & 4.9 & 4.7 & 4.5 & 6.1 & 12.3 \\
No-D        & 2.1 & 1.9 & 1.8 & 4.2 &  7.6 \\
Small-K     & 1.0 & 0.8 & 0.7 & 5.2 &  9.8 \\
Large-K     & 0.7 & 0.6 & 0.6 & 9.5 & 14.1 \\
Rand-Hook   & 2.5 & 2.3 & 2.0 & 4.4 & 11.5 \\
\textbf{Ours} & \textbf{0.4} & \textbf{0.3} & \textbf{0.3} & 5.8 & 10.4 \\ 
\bottomrule
\end{tabularx}
\label{tab:ablation}
\vspace{-0.5em}
\end{table}

Tab.~\ref{tab:ablation} shows that omitting condensation
(\textbf{No-C}) leaves \(>\!4\%\) true edges undiscovered because high
graph diameter prevents far-range stitching.  
Skipping densification (\textbf{No-D}) still misses nearly \(2\%\)
local edges despite perfect global connectivity.  
When the window is too small (\(K=c\)) the edit rate doubles, matching
the theoretical lower bound \(K\!\ge\!2(1+\varepsilon)c\); a very large
window (\(4c\)) brings only marginal accuracy gain at almost
double the probe budget.  
The random-hook variant confirms that \(s=\Theta(\log N)\) probes per
vertex are crucial for reliable connectivity.  
Our full pipeline keeps the edge-edit rate below \(0.4\%\) on 500–5 000
slices while maintaining \(\approx6\) oracle calls per section,
consistent with the \(\Theta(N(\log N+K))\) prediction.

\section{Trade-off between Field of View and Imaging Resolution and Seriation Accuracy}

In order to maximize the speed and accuracy of seriation, it is essential to select the optimal field of view (FOV) and image resolution. 

To systematically assess the trade-offs between computational efficiency and ordering robustness across various imaging configurations, we conducted extensive benchmarking on the H01 dataset, using 500-section sub-volumes sampled across a range of FOV and resolution (nm/px) settings. For each configuration, we recorded the total computation time, ordering outcome (success, swap error, or failure), and aggregated results into a two-dimensional heat map with log-scaled timing blocks.

\begin{figure}[H]
    \centering
    \includegraphics[width=1.0\linewidth]{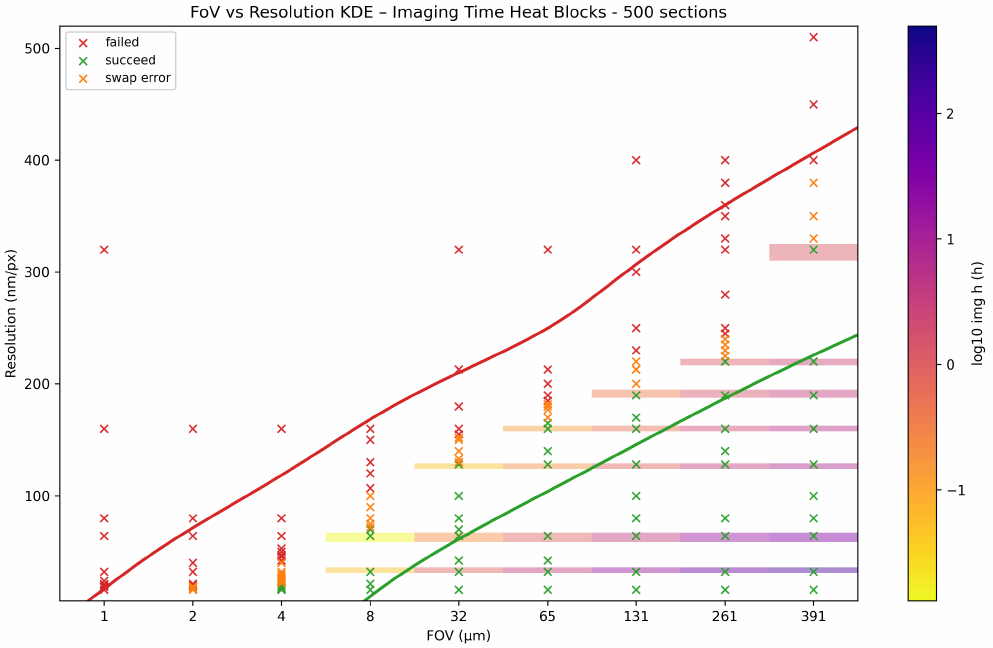}
    \caption{\textbf{FoV--Resolution landscape for a 500-section H01 stack.}   
    Markers indicate reconstruction outcomes: green (success), orange (swap error), and red (failure). 
    Colored heat blocks encode log-scaled computation times, with isocurves highlighting performance boundaries. 
    The visualization illustrates the operational regimes of success and failure under varying imaging configurations.
}
    \label{fig:Fov-res1}
\end{figure}

Our analysis reveals that FOV is the primary driver of computational cost, with larger fields of view incurring significantly longer runtimes. In contrast, high-resolution imaging exerts only a modest effect on computation time within our graph-based pipeline. This is because feature extraction in high-resolution images quickly reaches a saturation point, after which additional pixels do not proportionally increase the number or complexity of keypoints detected Fig.~\ref{fig:Fov-res1}.

Failures (red ×) concentrate at very small FOVs because in-plane translations between unsorted sections are unavoidable; even with \textsc{WaferTools}, which unifies ROIs and pre-targets acquisition before any alignment, residual shifts of a few microns persist \cite{wafertools2025}. This reduces effective overlap and makes local matches ambiguous, forcing reliance on long-range links that are noisier and can destabilize the global graph. Swap errors (orange x) cluster near the success/failure boundary, indicating transitional sensitivity. Successful orderings (green x) occur in the mid-range, roughly 32–131~\(\mu\)m FOV and 32–132~nm/px where overlap and runtime/accuracy are well balanced.

\begin{figure}[H]
    \centering
    \includegraphics[width=1.0\linewidth]{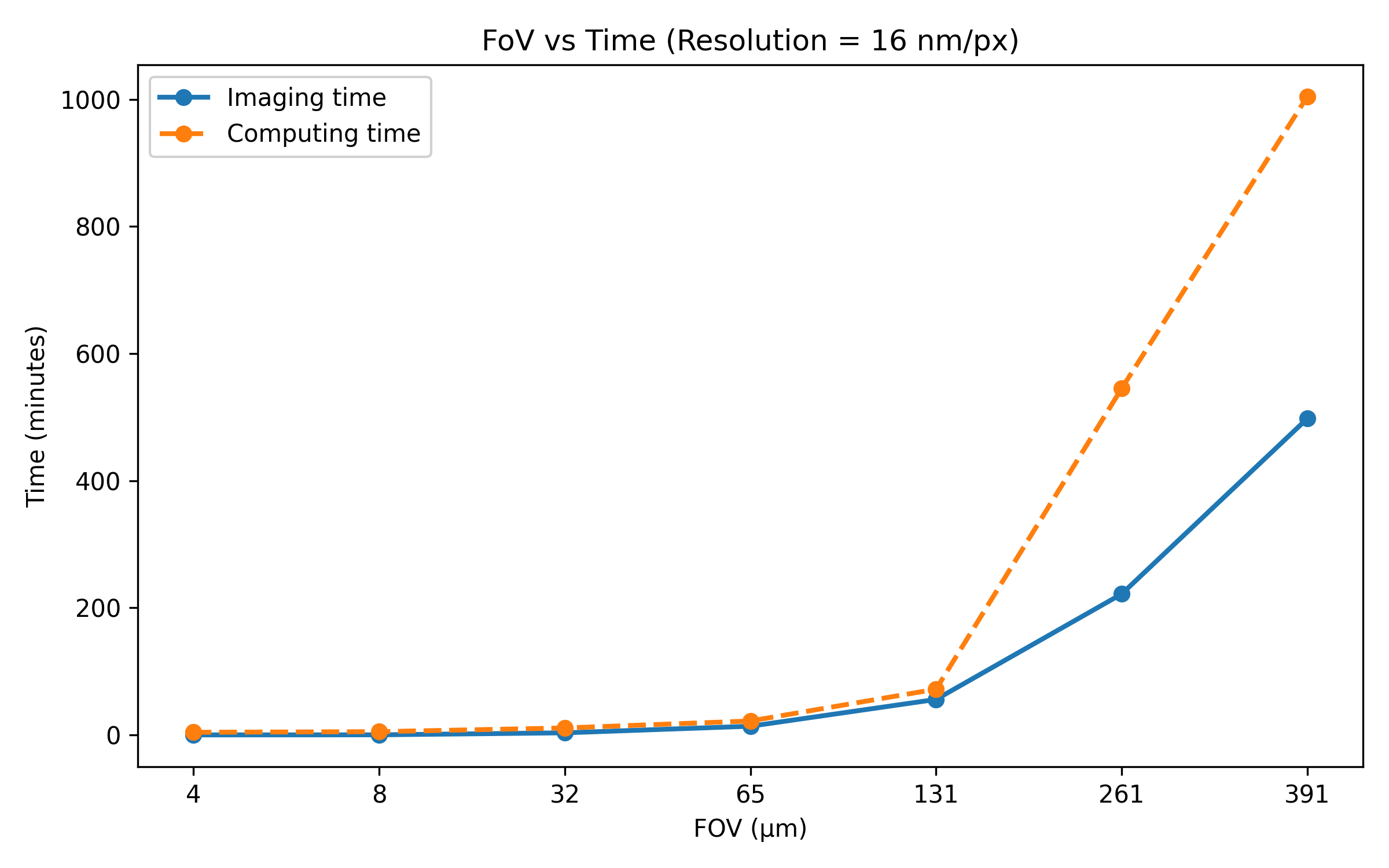}
    \caption{\textbf{Trade-off Between Imaging Time and Computational Burden Across Field-of-View.}   
 At a fixed 16\,nm/px resolution, computing time begins to exceed imaging time beyond $\sim$131\,µm FoV, highlighting a scalability bottleneck in naïve pairwise approaches.
}
    \label{fig:Fov-res2}
\end{figure}

We further compared imaging versus computational costs at a fixed resolution of 16\,nm/px (Fig.~\ref{fig:Fov-res2}). While imaging and computation remain comparable for small to moderate FoVs (4–65\,µm), beyond approximately 131\,µm the computational cost begins to exceed the imaging cost, scaling superlinearly with FoV due to the combinatorial explosion of pairwise matches. At the largest tested FoV (391\,µm), computation required nearly twice the imaging time ($>$1000 minutes for a 500-section stack), underscoring a fundamental scalability bottleneck.

These findings highlight a practical design constraint for large-scale connectomic acquisition: constraining FOV size is critical for computational tractability, while resolution may be flexibly optimized based on downstream biological needs without overwhelming system performance.

\section{Models and Theoretical Guarantees}
\label{sec:theory}

We provide a rigorous analysis under two oracle models: a binary oracle that indicates whether sections are within short range, and a noisy distance oracle that returns perturbed measurements. Our main results give near-linear query complexity and exact recovery guarantees under reasonable assumptions.

\subsection{Problem Setup and Notation}

Let $v_1, \ldots, v_N \in \mathbb{R}$ denote the unknown true positions of the $N$ sections, with the ground-truth permutation $\pi^*$ sorting indices by increasing position: $v_{\pi^*(1)} < \cdots < v_{\pi^*(N)}$. We adopt unit normalization where the minimum gap between adjacent sections is at least $1$. The notation used throughout our analysis is summarized in Table~\ref{tab:notation}.

\begin{table}[t]
\footnotesize
\setlength{\tabcolsep}{4pt}
\centering
\begin{tabularx}{\linewidth}{@{}lX@{}}
\toprule
\textbf{Symbol} & \textbf{Meaning} \\ \midrule
$N$                              & number of sections / vertices \\[0.2em]
$\rho\ (>0)$                     & reliable short-range radius in the weighted oracle \\[0.2em]
$c = \lceil \rho \rceil$         & short-edge radius in index distance \\[0.2em]
$\varepsilon$                    & relative noise level in the weighted oracle ($0 \leq \varepsilon < 1$) \\[0.2em]
$\delta = (1-\varepsilon)\rho$   & acceptance threshold on observed distance \\[0.2em]
$K$                              & densification window width (default $K=\gamma c$ with $\gamma\!\ge\!4$) \\[0.2em]
$\gamma = K/c$                   & window-to-radius ratio (default $\gamma\!=\!4$) \\[0.2em]
$s = \Theta(\log N)$             & random probes per vertex in Phase 1 \\[0.2em]
$\Delta$                         & minimum weight gap between true and false edges \\[0.2em]
$\mathrm{diam}(G)$               & diameter of graph $G$ \\[0.2em]
$\mathrm{dist}_G(u,v)$           & graph distance between vertices $u$ and $v$ \\
\bottomrule
\end{tabularx}
\caption{Key notation used in theoretical analysis.}
\label{tab:notation}
\end{table}

\subsection{Binary Oracle Model}

In the binary oracle model, we can query whether two sections are within a short-range radius $c \geq 2$:
\begin{equation}
E(i,j) \;=\; \mathbf{1}\!\bigl[\,|v_i - v_j| \leq c\,\bigr].
\end{equation}
This induces an unknown graph $G^* = (V, E^*)$ where edges exist between all pairs within distance $c$. Our algorithm discovers a subset of these edges sufficient for exact recovery.

\begin{lemma}[Diameter Recurrence]
\label{lem:diameter_recur}
Let $G$ be any connected spanning subgraph of $G^*$. A condensation round augments $G$ by adding all oracle-positive edges incident to a small set of farthest vertices (chosen by a double-sweep on $G$). The resulting graph $G'$ satisfies
\[
\mathrm{diam}(G') \;\le\; \Big\lceil \tfrac{\mathrm{diam}(G)}{2} \Big\rceil + c .
\]
\end{lemma}

\begin{theorem}[Bounded Diameter After Condensation]
\label{thm:bounded_diameter}
Starting with any spanning tree $T$ from Phase 1, after $t = \lceil\log_2 \mathrm{diam}(T)\rceil$ rounds of condensation we obtain
\[
\mathrm{diam}(G_t) \;\le\; \frac{\mathrm{diam}(T)}{2^t} + 2c \;\le\; 2c + 1 .
\]
\end{theorem}

\begin{theorem}[Query Complexity — Binary Oracle]
\label{thm:binary_main}
Fix $K=\gamma c$ with a constant $\gamma\ge 4$. There exists an algorithm that recovers the true permutation $\pi^*$ using
\[
O\!\bigl(N(\log N + K)\bigr)
\]
oracle queries with probability $1 - O(N^{-1})$: $O(N\log N)$ in Phases 1–2 and at most $2NK$ candidate probes in the $K$-window densification. The final ordering is obtained by a chaining routine (\textsc{SuperChain}) in $O(N\log N)$ time.
\end{theorem}

\subsection{Noisy Distance Oracle Model}

In practice, similarity measurements are corrupted by noise. We model this with an oracle that returns
\begin{equation}
D(u,v) \;=\; |v_u - v_v| + \eta_{uv},
\label{eq:noisy_oracle}
\end{equation}
and assume the following global one-sided lower bound:
\[
\textbf{(W1)}\qquad D(u,v)\;\ge\;(1-\varepsilon)\,|v_u-v_v|\quad \text{for all } u,v .
\]
Let $\rho>0$ be a short-range reliability radius and set $c:=\lceil\rho\rceil$. We accept an edge iff $D(u,v)\le \delta=(1-\varepsilon)\rho$.

\begin{lemma}[Edge Validity Under Noise]
\label{lem:edge_validity}
Under \textbf{(W1)}, if $D(u,v) \le \delta$, then $|v_u - v_v| \le \rho$, so $(u,v)$ connects vertices at most $c=\lceil \rho\rceil$ positions apart in the true order.
\end{lemma}

\begin{lemma}[Position Error After Condensation]
\label{lem:position_error}
Let $g_r = \max_{v \in V} |\hat{\pi}_r^{-1}(v) - \pi^{*-1}(v)|$ denote the maximum rank error after $r$ condensation rounds. Then
\[
g_{r+1} \;\le\; \frac{g_r}{2} + \lceil \rho \rceil .
\]
\end{lemma}

\begin{theorem}[Window Sufficiency]
\label{thm:window_sufficient}
Choose $K \ge 4\lceil\rho\rceil$ and
\(
t \;\ge\; \left\lceil \log_2\!\frac{g_0}{\,K/2-2\lceil\rho\rceil\,}\right\rceil.
\)
Then $g_t \le K/2$, so every true adjacent pair lies within distance $\le K$ in the provisional ordering and will be probed during $K$-window densification.
\end{theorem}

\begin{theorem}[Main Result — Weighted Oracle]
\label{thm:weighted_main}
With probability $1 - O(N^{-1})$, the four-phase algorithm issues
\[
O\!\bigl(N(\log N + K)\bigr)
\]
oracle queries and outputs the exact permutation, provided $K \ge 4\lceil\rho\rceil$ and the acceptance threshold is $\delta=(1-\varepsilon)\rho$.
\end{theorem}
\subsection{Exact Recovery of Order Under SuperChain}
\label{sec:superchain}

After the sparse graph reconstruction, we recover the order with a simple
two-ended greedy chaining procedure.

\textbf{Assumptions.}
Let $\pi^\star=(v_1,\dots,v_N)$ be the ground-truth order. We assume:
\begin{enumerate}[label=\textbf{A\arabic*.},leftmargin=1.9em,itemsep=0pt]
  \item \emph{Top-1 correctness.}
        For every internal vertex $v_i$ ($2\le i\le N-1$), its two true incident edges
        $e_i^-=(v_{i-1},v_i)$ and $e_i^+=(v_i,v_{i+1})$ are present and maximize $w$ at $v_i$.
  \item \emph{Tie-free margin.}
        There exists $\Delta>0$ such that for any non-true edge $(v_i,u)\in E$,
        $w(e_i^\pm)\ge w(v_i,u)+\Delta$.
  \item \emph{Endpoint purity.}
        The globally heaviest edge $e^\star=\arg\max_{e\in E}w(e)$ is a true edge
        $(v_s,v_{s+1})$ (implied by A1).
\end{enumerate}


\begin{lemma}[Endpoint local maximality]
\label{lem:endpoint-max}
Under A1–A2, let $C=(v_i,\dots,v_j)$ be any contiguous substring of $\pi^\star$.
At endpoint $v_i$ (resp.\ $v_j$), the heaviest unused incident edge is uniquely
the true boundary edge $(v_{i-1},v_i)$ (resp.\ $(v_j,v_{j+1})$).
\end{lemma}

\begin{proof}
At $v_i$, by A1 the two true incident edges are the top weights at $v_i$.
One of them, $(v_i,v_{i+1})$, is already used inside $C$; the other,
$(v_{i-1},v_i)$, is unused and, by A2, strictly heavier than any non-true
unused edge. Uniqueness follows from the $\Delta$ margin. The right endpoint
is symmetric.
\end{proof}

\begin{theorem}[Exact recovery under \textsc{SuperChain}]
\label{thm:superchain}
Under A1–A3, the above procedure outputs $\pi^\star$ in $N-1$ iterations
and $O(N\log N)$ time.
\end{theorem}

\begin{proof}
\emph{Contiguity \& progress in one step.}
Suppose $C_t$ is contiguous. By Lemma~\ref{lem:endpoint-max}, the algorithm
can only append $v_{i-1}$ or $v_{j+1}$, hence $C_{t+1}$ remains contiguous
and gains exactly one new vertex. Induction from $C_0=[v_s,v_{s+1}]$ (A3)
gives $|C_{N-1}|=N$ and $C_{N-1}=\pi^\star$.

For each endpoint, a max-heap over its incident unused edges.
Each vertex is appended once and each edge is touched $O(1)$ times, giving
$O(N\log K)$, hence $O(N\log N)$.
\end{proof}

When top-2 neighborhoods are not always correct, a sufficient and checkable condition is: at least three inter-chain gaps (between the worst-case $N/2$ length-2 mini-chains induced by top-1 links) are fully supported (both endpoints rank each other within top-2); then the same chaining recovers $\pi^\star$ deterministically.
Empirically, recovery typically occurs once the overall fraction of
second-best-correct vertices approaches $\sim\!2/3$ (fig:~\ref{fig:recovery_probability}).

\section{Conclusion}

We presented a near-linear solution to the large-scale section ordering problem in connectomics. Our approach combines five key components: (1) random-hook Borůvka for initial connectivity, (2) iterative graph condensation to bound diameter, (3) a double-sweep BFS to obtain a provisional global order, (4) fixed-window densification for local edge recovery, and (5) \textsc{SuperChain} merging that exploits empirical top-1 + $\Delta$ margins. This pipeline reduces computational complexity from $O(N^2)$ to $\Theta\!\bigl(N(\log N+K)\bigr)$ while maintaining perfect accuracy on all tested datasets. The method enables overnight processing of thousands of sections on commodity hardware, removing a critical bottleneck in whole-brain connectome reconstruction.

\section{Acknowledgments}
This work was supported by the National Institutes of Health (NIH) under award UM1NS132250. 

We thank members of the Lichtman Lab for their support. We are especially grateful to Richard Schalek for sectioning the HPF sample block; to Vlad Susoy for project support and help with revising the manuscript; and to Neha Karlupia for preparing the HPF sample block.

%

{
    \small
    \bibliographystyle{ieeenat_fullname}
    \bibliography{main}
}

\end{document}